# The Relativistic Transactional Interpretation and Spacetime Emergence

R. E. Kastner*
3/10/2021

*"The flow of time is a real becoming in which potentiality is transformed into actuality."* -Hans Reichenbach[1]

Abstract. We consider the manner in which the spacetime manifold emerges from a quantum substratum through the transactional process, in which spacetime events and their connections are established. In this account, there is no background spacetime as is generally assumed in physical theorizing. Instead, the usual notion of a background spacetime is replaced by the quantum substratum, comprising quantum systems with nonvanishing rest mass. Rest mass corresponds to internal periodicities that function as internal clocks defining proper times, and in turn, inertial frames that are not themselves aspects of the spacetime manifold, but are pre-spacetime reference structures. Specific processes in the quantum substratum serve to distinguish absolute from relative motion.

## 1. Introduction and Background

The Transactional Interpretation of quantum mechanics has been extended into the relativistic domain by the present author. That model is now known as the Relativistic Transactional Interpretation (RTI) (e.g., Kastner 2015, 2018, 2019, 2021; Kastner and Cramer, 2018). RTI is based on the direct-action theory of fields in its relativistic quantum form, as presented by Davies (1971, 72). The current paper focuses only on the issue of spacetime emergence. Readers desiring fully relativistic details of the quantum-level transactional process may consult Kastner (2021). Briefly, RTI involves two kinds of dynamical processes: (1) the usual unitary evolution as reflected in the time-dependent Schrödinger equation (or Heisenberg evolution for operators in the Heisenberg picture), and (2) a non-unitary process corresponding to the formation of a transaction and the transfer of conserved physical quantities. We will refer to these processes as the U-interaction and the NU-interaction, respectively. Standard approaches to quantum theory acknowledge only the U-interaction, which describes the influence of forces but not real energy transfer (this is arguably why the standard theory has a measurement problem). One obtains (2), the NU-interaction, only in the direct-action theory of fields (which nevertheless is empirically equivalent to the standard theory at the level of Born probabilities). Kastner (2021) discusses this issue in detail. It is the NU-interaction that gives rise to the emergence of determinate spacetime events that is the subject of this paper.

---

* rkastner@umd.edu; Foundations of Physics Group, University of Maryland, College Park

[1] Reichenbach, H. (1953). (Trans.: M. Čapek.)

The ontology of spacetime in the Relativistic Transactional Interpretation (RTI) differs quite radically from the usual conception of spacetime that (often uncritically) underlies theorizing in physics. In particular, RTI does not assume the usual notion of a "spacetime background" that is either implied or explicitly invoked in physical theories. In what follows, we will consider in quantitative terms the manner in which a structured set of events constituting "spacetime" emerges from the quantum level by way of actualized transactions, in a process of genuine "becoming." Thus, the spacetime manifold emerges from the extra-spatiotemporal *quantum substratum* comprising physical *potentiae*; i.e., entities described by quantum states. This domain is characterized by Hilbert space structures and processes (such as unitary interactions).

We must first recall from Kastner (2019) and (2021) that the only kind of quantum system that is transferred directly from an emitter to an absorber in an actualized transaction is the massless gauge boson, or photon.[2] Quantum systems with non-vanishing rest mass act as emitters and absorbers of photons, and are not themselves transferred in transactions, i.e, via offers and confirmations. While they can be transferred from one system to another, such transfers involve liberation from one bound state and re-integration into another bound state, a process that is physically distinct from the emission and absorption of a photon in a transaction. In what follows, we will see that the role of emitters and absorbers in spacetime emergence is quite different from that of the photon. We'll also see how transactions create a discrete, interlocking, structured set of events that fulfills the role of "spacetime" without the necessity of invoking a background spacetime "substance" or container. Thus, according to RTI, the usual notion of a "spacetime continuum" is a fiction, as is the notion of a "spacetime background" for all physical processes.

It turns out that a natural way to formulate the process of spacetime emergence is in terms of the concept of a *causal set*. We will begin by reviewing this concept.

## 2. Causal set approach to spacetime

A causal set is a partially ordered set that can enlarge in a directed manner. The primary motivation for the causal set program was to solve the problem of quantum gravity. Its originator, Raphael Sorkin, remarked:

> The causal set idea is, in essence, nothing more than an attempt to combine the
> twin ideas of discreteness and order to produce a structure on which a theory of
> quantum gravity can be based. That such a step was almost inevitable is indicated by the fact
> that very similar formulations were put forward independently in [G. 't Hooft
> (1979), J. Myrheim (1978) and L. Bombelli et al (1987)], after having been adumbrated

---

[2] Thus far, gluons are considered massless, but their status in this regard is uncertain. In any case they are never free particles in the sense that they do not carry radiative energy, so we disregard them as far as transactions are concerned.

in [D. Finkelstein (1969)]. The insight underlying these proposals is that, in passing from the continuous to the discrete, one actually gains certain information, because "volume" can now be assessed (as Riemann said) by counting; and with both order and volume information present, we have enough to recover geometry. (Sorkin 2003, p. 5)

While the transactional interpretation is not a theory of quantum gravity, it dovetails very naturally with the above program in that the structures that emerge from the transactional process feature both discreteness and order, and effectively form a causal set.[3]

In formal terms, a causal set C is a finite, partially ordered set whose elements are subject to a binary relation $\prec$ that can be understood as precedence; the element on the left precedes that on the right. It has the following properties:

(i) transitivity: $(\forall x, y, z \in C)(x \prec y \prec z \Rightarrow x \prec z)$
(ii) irreflexivity: $(\forall x \in C)(x \not\prec x)$
(iii) local finiteness: $(\forall x, z \in C)$ (cardinality $\{ y \in C \mid x \prec y \prec z \} < \infty$)

Properties (i) and (ii) assure that the set is acyclic, while (iii) assures that the set is discrete. These properties yield a directed structure that corresponds well to temporal becoming, which Sorkin describes as follows:

> the relationship $x \prec y$ … is variously described by saying that x precedes y, that x is an ancestor of y, that y is a descendant of x, or that x lies to the past of y (or y to the future of x). Similarly, if x is an immediate ancestor of y (meaning that there exists no intervening z such that $x \prec z \prec y$) then one says that x is a parent of y, or y a child of x,
> …or that $x \prec y$ is a link. (Sorkin 2003, p. 7)

In Sorkin's construct, one can then have a totally ordered subset of connected links (as defined above), constituting a *chain*. In the transactional process, we naturally get a parent/child relationship with every transaction, which defines a link. Each actualized transaction establishes three things: the emission event *E*, the absorption event *A*, and the invariant interval *I(E,A)* between them, which is defined by the transferred photon. Thus, the interval *I(E,A)* corresponds to a link. Since it is a photon that is transferred, every actualized transaction establishes a null interval, i.e., $ds^2 = c^2t^2 - r^2 = 0$. The emission event *E* is the parent of the absorption event *A* (and *A* is the child of *E*).

A major advantage of the causal set approach as proposed by Sorkin and collaborators (e.g., L. Bombelli, Meyer and Sorkin, 1987) is that it provides a fully covariant model of a growing spacetime. It is thus a counterexample to the usual claim (mentioned in the previous section) that a growing spacetime must violate Lorentz covariance. Specifically, Sorkin shows that if the events are added in a Poissonian manner, then no preferred frame emerges, and covariance is preserved (Sorkin 2003, p. 9).

---

[3] Nevertheless, the ability of RTI to provide an account of the emergence of elements of the causal set positions it as a useful component of a theory of quantum gravity of the sort Sorkin *et al* are exploring.

In RTI, events are naturally added in a Poissonian manner, because transactions are fundamentally governed by decay rates (Kastner and Cramer, 2018). As discussed in Kastner (2018) and (2021), the elementary probability of the NU-interaction (occurrence of a transaction) corresponds to the fine structure constant $\alpha$. But, owing to conservation requirements, the full expression for the probability of a transaction is essentially the transition probability between emitter/absorber states, $X$ (excited) and $G$ (unexcited): $|\langle X,0|H_{int}|G,k\rangle|^2$. Here, $H_{int}$ is the interaction Hamiltonian quantifying the coupling between the emitting/absorbing charges and the electromagnetic field: $H_{int} = e\hat{A}\cdot\vec{p}$ in natural units.[4] $k$ is the state of the photon that is transferred in order to satisfy conservation requirements.

The squared form emerges in the transactional picture because both emission and absorption are necessary for transfer of the photon, and the photon emission and absorption amplitudes are complex conjugates of one another. When one multiplies both amplitudes together for the complete process, one therefore gets the Born probability of a photon being transferred between these two states. In the squaring of a transition amplitude containing $H_{int}$ we see the origin of the factor of $\alpha=e^2$, the fine structure constant. Applying time-dependent perturbation theory to the specific initial and final states, given the perturbation $H_{int}$, leads to the standard decay rate, a Poissonian process.[5] Thus, the emergent spacetime structure in RTI is fully covariant. However, it's important to note that, while the original causal set model assumes that individual events constitute the basic volume element of spacetime, in RTI it is the invariant spacetime interval $I(E,A)$ that constitutes the basic volume element. So, rather than a structure that is growing by single events, the RTI spacetime structure grows by pairs of connected photon emission and absorption events—that is, by links. As noted above, this means that the spacetime manifold itself is really constituted solely of null intervals.

## 3. Rest mass remains in the quantum substratum

Before studying the resulting structure further, we must be clear about an unfamiliar aspect of the proposed ontology. This ontology departs sharply from the usual Democritan concept of "atoms in the void" upon which physics has been traditionally based (however unconsciously at times). That is, it is usually tacitly assumed that physics deals with chunks of something called "matter" moving around in an otherwise empty spacetime container. Matter, in this picture, is an undefined primitive with only operational properties.

However, in the RTI ontology, systems with rest mass, such as atoms and molecules—i.e., emitters or absorbers—are *not* part of the spacetime manifold. They remain in the extra-spatiotemporal domain (quantum substratum) described by Hilbert space, even as they undergo state changes as a result of their participation in transactions. This means that the notion of "change" applies just as well to the quantum substratum as it does to the spacetime manifold. In Section 5, we'll see that quantum-level change can be described by reference to an internal "clock" of quantum systems with finite rest mass.

---

[4] The direct-action theory can work with the Hamiltonian form because of the equivalence of the traditional quantum field $\hat{A}$ with a direct connection between currents, as discussed in Kastner (2021).

[5] Here, the perturbation applies to the interacting fermions, which evolve according to the time-dependent Schrödinger equation. The squaring, strictly speaking, really applies to the photon, which is not tied to any spatiotemporal index (unlike the emitter and absorber for whom inertial frames can be defined). This is related to the fact that emitters and absorbers are detected only indirectly by way of photon transactions, and are not themselves transferred via transactions.

Emission and absorption *events* such as *E* and *A* above must be distinguished from the emitter and absorbers themselves; an event is *not* a rest-mass quantum system. An event is not an entity or substance; rather, it is an *activity* of an entity. The only elements of spacetime are emission and absorption events and the real photon defining and connecting the two events. Thus, emission/absorption events and the transferred photons (constituting links) are aspects of spacetime structure; everything else is not, and abides in the quantum substratum.[6] This substratum is the source of the emerging, growing spacetime structure, much as the mineral-laden water is the source of the crystals in a geode. The disanalogy here is that rest-mass quanta do not themselves transform into spacetime objects; only the electromagnetic field does so, in the form of photons that serve as structural elements (links) of the emergent spacetime manifold.

This picture has much in common with Ellis and Rothman's "Crystallizing Block Universe" (CBU), which is a particular sort of "growing spacetime" (Ellis and Rothman, 2010). However, the CBU ontology seems to assume a spacetime background, even if the future is taken as indefinite. Thus, RTI differs somewhat from the CBU picture in that in RTI the quantum formalism specifically refers to an extra-spatiotemporal domain, or quantum substratum, from which the spacetime structure emerges. In addition, the actualization of events in RTI does not correspond to a "moving present" that progresses "toward the future" as is the case in the CBU. Instead, as discussed in the previous section, the generated spacetime structure *recedes* from a present that is eternal in some sense, since it is just the interface between the quantum level of possibilities and the growing spacetime manifold. Again, the knitting analogy can be helpful here, although it should be kept in mind that there is no preferred reference frame corresponding to a "knitting needle." In this respect, it agrees with the CBU picture in that it is interacting matter and energy that locally generates actualized events, in a manner that does not single out any preferred reference frame.

In addition, under RTI, the actualization of measurement results corresponds to a specific quantitative physical process (i.e., the transactional process or NU-interaction), and it is not dependent on decoherence arguments or top-down considerations, as is the case with the CBU. Rather, the actualization of spacetime events and the attendant arrow of time emerge from the micro-level, through the non-unitarity of the NU-interaction. This issue was discussed in Kastner (2017).

---

[6] Besides nonlocality and entanglement, another aspect of the departure of quantum-level processes from longstanding empirical-level physical principles is found in the fact, as noted in Brown (2002), that quantum test particles do not obey the "zeroth law of mechanics," i.e., the principle that "the behavior of free bodies does not depend on their mass and internal composition" (Brown, 2002, p. 25). This is easily seen by looking at the time-dependent Schrödinger equation, which depends explicitly on the mass of the quantum. Again, this discrepancy can be understood by considering quantum mechanics as describing the behavior of sub-empirical (pre-spacetime) objects that do not have to obey empirical-level principles of mechanics.

## 4. The basic structure of the emergent spacetime manifold

With the above in mind, let us recall the parent/child relationship introduced in the previous section, and consider a single actualized transaction involving an emitter, C, and its receiving absorber, D. (Refer to Figure 1.) C and D are bound systems such as atoms or molecules, i.e., systems with internal degrees of freedom subject to excitation. D is the absorber that actually receives the real photon as a result of the final collapse (or reduction).

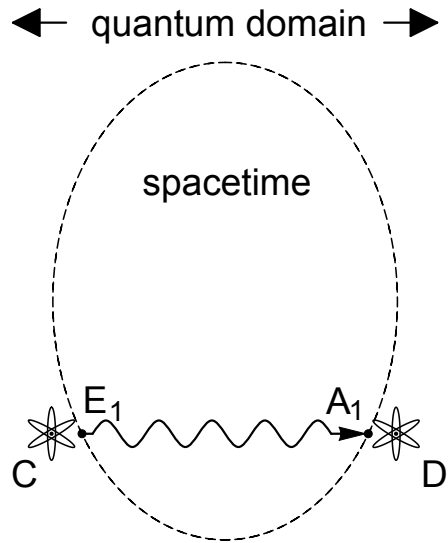

Figure 1. A single transactional link is created.

The atoms' initial roles as emitter and absorber can be represented by denoting their initial states as $|X\rangle_C$ and $|G\rangle_D$, respectively, where $X$ is the excited state and $G$ the ground state. Let us designate the initial emission event as $E_1$. This event heralds C's transition from the excited state to the ground state, $|X\rangle_C \rightarrow |G\rangle_C$. Similarly, the absorption event $A_1$ heralds D's transition from the ground state to the excited state, $|G\rangle_D \rightarrow |X\rangle_D$. The newly created link is a null interval established by the transferred photon (indicated by the wavy line) and bounded by $E_1$ and $A_1$: symbolically, $I(E_1,A_1)$. The figure schematically depicts the idea that the emitter and absorber remain beyond the spacetime construct, in the quantum substratum, while the exchanged photon establishes a link that constitutes an element of spacetime. In addition, the bold dots indicate the spacetime events of the emission and absorption (again, these events are not identified with the emitting and absorbing systems but rather are activities of those systems).

Following the absorption event $A_1$, D is now in the excited state $|X\rangle_D$ and is therefore poised to become an emitter that could emit to a new absorber F, or back to the original emitter C, which, now in its ground state, $|G\rangle_C$ serves as a potential absorber (see Figure 2). However, there is a gap between D's absorption event $A_1$ and D's subsequent emission event $E_2$, because these are distinct events; D plays a different role in each. This reveals that spacetime is not only discrete, but is also a discontinuous structure, in the sense that it consists of independent and distinct photon emission-and-absorption "links." In this respect, the RTI picture differs from the causal set picture in that the chains are not continuous. Figure 2 shows several transactional links, and the gaps between them, where the latter involve the continued existence of the participating atoms in the quantum substratum.

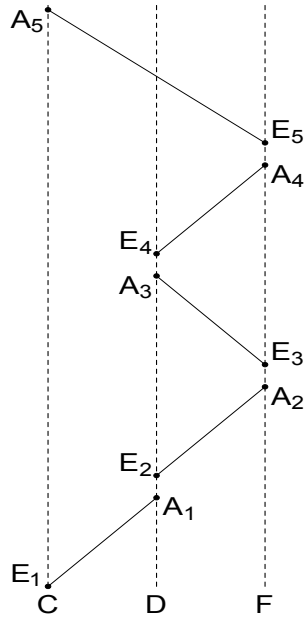

Figure 2. Sequential emissions and absorptions.

We now come to an interesting point. Figure 2 shows what appears to be a spacetime diagram; however, since the atoms C,D, and F are never "in spacetime," their "temporal axes" are not really properties of spacetime. Instead, they are *internal* references only. We can think of these as reflecting the counting of internal clocks, i.e., strictly local periodic processes that are sequential in nature, such that they yield a locally increasing index. This locally increasing index corresponds to the time coordinate of the atom's rest frame—its proper time. But what is seemingly paradoxical is that this "time coordinate" is not really a property of the spacetime construct! It is just an *internal* reference that is used to index spacetime events from a particular reference frame—the rest frame of the quantum system (such as electron or atom) doing the indexing. And this is why it is not an invariant quantity. Thus, we arrive at the following picture: inertial frames are *internal reference structures* of entities in the quantum substratum, *not* aspects of spacetime itself. *Inertia comes from the quantum substratum!* This idea is reinforced by the fact that real photons, which are part of the spacetime construct, do not possess inertial frames.

Spacetime itself is constructed only of invariant quantities: events themselves, and the spacetime intervals or links established by real photons. The irony here is that even though we call this manifold "spacetime," it is not constructed of "space" and "time." These are just frame-dependent parameters used as labels for the connected *events* that actually comprise spacetime. And what connects these events is momentum and energy—really, 4-momentum, as contained in the transferred photon. It is in this sense that four-momentum generates spacetime displacements (but only relative to a given inertial reference frame). Energy transfer corresponds to temporal displacement, while 3-momentum corresponds to spatial displacement. But "temporal displacement" and "spatial displacement" are *not* themselves aspects of the "spacetime" construct. Both are merely non-invariant descriptions relative to a given inertial frame. The inertial rest frame and attendant "proper time" is defined by the quantum system's rest mass—the invariant quantity and the source of the system's inertia and attendant rest frame.

We can get an idea of what might constitute a physical internal clock corresponding to rest mass by reference to the deBroglie frequency,

$$\omega_{DeB} = \frac{m_e c^2}{\hbar} \qquad (1)$$

David Hestenes has constructed a useful model of the electron by incorporating "Zitterbewegung," i.e., the fundamental oscillatory motion that is the source of electron spin (e.g., Hestenes 2010, 2019). In this model, the origin of the electron's rest mass is the energy associated with its motion in a lightlike helix (see Figure 3).

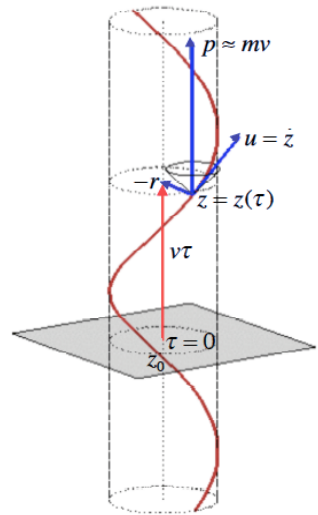

Figure 3. The "Zitter" model of the electron by D. Hestenes. Used with permission.[7]

---

[7] Hestenes, D. (2010).

This helical motion defines a timelike "world tube" corresponding to an effective sub-luminal rectilinear momentum. The model provides correct correspondence with the Dirac theory of the electron. In particular, it obtains a fundamental angular frequency that modifies the deBroglie frequency with a factor of ½, i.e.: [8]

$$\omega_e = \frac{2m_e c^2}{\hbar} \qquad (2)$$

The corresponding periodic "clock" function can then be written as

$$\psi(\tau) = e^{i\omega_e \tau}, \qquad (3)$$

where $\tau$ is the electron's proper time. However, it's important to note that this oscillatory "motion" is *not* spacetime motion. The electron's half-integral spin involves a nontrivial topology, and we cannot pretend that what is "revolving" is a classical point particle. Rather, the rotational motion is that of a spinor—a specific quantum entity that undergoes a sign change upon a rotation of $2\pi$. This means that a "double rotation" of $4\pi$ is required for return of the electron to its initial state.[9] In any case, the fundamental frequency in (2) provides a periodic process that serves as the "internal clock" that provides the proper time reference for the electron; i.e., that defines the temporal axis of its rest frame, and which therefore serves to define the concept of an inertial frame. The concept of an inertial frame is physically distinguished from accelerating frames via the fact that the fundamental frequency (internal clock) is obtained from the Dirac equation for a free (non-interacting) electron.

We digress here briefly to note that the foregoing resolves a longstanding puzzle regarding whether Newton's concept of "absolute space" is needed to define non-inertial motion such as that of the rotating water in Newton's famous spinning bucket experiment. We now see that both forms of motion—inertial and non-inertial—are defined in terms of the quantum substratum, not "absolute" space, time, or a substantive "spacetime continuum." Specifically, inertial motion is defined by reference to non-interacting fermionic systems (or bound systems composed of fermionic fields), while non-inertial motion is defined by reference to interacting fermionic systems. *Thus, it is the quantum substratum that constitutes the absolute reference for types of motion.*

Similar considerations apply to any quantum system with nonvanishing rest mass, such as larger fermions, atoms and molecules. It is rest mass that generates the internal clock that serves to define the proper time and thus the temporal axis for the system. Metaphorically speaking, we can think of rest mass as the "sand in the hourglass." Rest mass originates in the nontrivial topology of fermions—quantum systems with half-integral spin. We can think of fermions as forms of "trapped light," since the fundamental structure is that of an electromagnetic field confined to a pre-spatiotemporal topological vortex. As discussed in Kastner, Kauffman and Epperson (2018), this quantum substratum is a form of physical possibility, so electrons and other quanta with rest mass retain the ability to enter into superpositions. However, they can be "collapsed" into determinate states through participation in transactions. The latter issue is quantitatively addressed in Kastner (2020).

---

[8] Schrodinger originally obtained this frequency by analyzing the time dependence of the velocity operator for the Dirac equation.

[9] In addition, Hestenes (2019) notes that the Dirac wavefunction really describes an ensemble of helices, not a single helix, though he proposes that the electron itself corresponds to one of the helices. In the present interpretation, the electron corresponds to the entire ensemble and should not be thought of as a point particle. This is because the idea of a position continuum is an idealization, and localization is only a phenomenon arising from the tiny (but finite) size of micro-absorbers such as atoms and molecules.

Let us return now to photons, having zero rest mass. Of course, the photon has no rest frame and is not an inertial object. From the vantage point of a photon, no time elapses between its emission and its absorption: the photon's "clock" is static. Neither is there any spatial separation, from the vantage point of a photon, between its emission event and its absorption event. For the photon, there is no distinction between the "time axis" and the "spatial axes"—they are merged as the photon's null internal. The distinction between time and space appears *only* by reference to an object with rest mass that defines an inertial frame. And "rest mass" is, in effect, electromagnetic energy confined to a topological vortex in the quantum substratum.

We noted above that links are established via photon transfers; the emission event is the parent of the absorption event. Considering again Figure 2, what about the timelike gap between events $A_1$ and $E_2$, in which atom D stands ready to emit after having absorbed? In a chronological sense, $A_1$ is the parent of $E_2$, but they are not connected by a transferred photon "link." We can describe this situation with the concept of an *implicit link,* or IL for short. An IL is not part of the spacetime construct or causal set, but it still contains temporal information, including an arrow of time, by reference to the inertial frame defined by the rest mass "clock" of the quantum system in question. Again, in this picture, inertial frames are *not* aspects of the spacetime manifold; they are internal, quantum-level references. Note that this reflects the vital, physical role played by the quantum substratum as the generator of spacetime, both in the active sense (via transactions that create spacetime links) and in the passive sense (by defining inertial frames).

## 5. Spacetime as an "influence network"

Another proposal for a causal set structure has been offered by Kevin Knuth and collaborators (e.g., Knuth and Bahreyni, 2012). Knuth *et al* champion a non-substantival, relational view of spacetime, and in that respect their approach has much in common with the RTI picture. They note the prevalence of the usual notion of spacetime as a fundamental, physical container, but then go on to say:

> However, more recently, the idea that space-time is neither physical nor fundamental has been growing [Seiberg, N. (2007)]. The idea is that space and time may emerge from more fundamental relations or phenomena … In addition to the older ideas, such as space as a container or space as a substance, which have mostly dominated our perspectives of space, is the view that space represents a relation between objects.[10]

These authors call their structure a "poset," for "partially ordered set." They assume, like the present author, that the only real elements of spacetime are events and their influences, which map to emission/absorption events and photon transfers in RTI respectively. Using only the assumption that an observer can be identified with each chain (totally ordered set of events), together with a consistency condition between observers and the "radar formulas" for timelike and spacelike displacements between events, they obtain the Minkowski metric. While this formulation is thus far restricted to 1+1 spacetime, it may be possible to extend it to 3+1.

---

[10] Knuth and Bahreyni (2012). p. 1, preprint version.

The RTI ontology differs in some respects from Knuth's poset picture (henceforth abbreviated "KPP"). As noted in the previous section, in RTI, objects with rest mass possess internal periodic "clocks," which serve to establish a proper time independently of specific events. While influences are a primitive concept in the KPP ontology, RTI physically specifies the nature of the influences in terms of photon transfers. In addition, in RTI influences are always mutual, since the transferred photon affects both the emitter and the absorber. This contrasts with KPP, in which an observer either influences or is influenced by another observer. While KPP assumes that determinate (classically describable) structures are revealed epistemically, from coarse-graining (using a scale that shows less detail), in RTI these emerge at an ontological level, from the non-unitary transactional process that transforms possibilities into actualities. Nevertheless, the approaches have much in common in that spacetime is a secondary, emergent construct that is fundamentally based on relations and interactions between quantum systems. KPP has great promise in that it manages to extract a great deal of information concerning the spacetime structure, including the Minkowski metric, merely by demanding consistency among the chains concerning the sharing of influences and attendant information about events. A step has also been taken towards accommodating general relativity in KPP, by quantifying the effects of mass on the poset structure. The result is an equation of geodesic form (Walsh and Knuth, 2015).

## 6. A Common Worry and Why It's not a Problem

In this section, I address a question that pops up from time to time as a possible objection to the transactional picture. The scenario involves a very distant star that engages in a transaction with a person's eye, so that they see the star as it existed billions of years ago. But suppose the star has long since ceased to exist, and that it sent out that photon long before this observer was born? How did the star "know" that the observer would be in the right place at the right time to engage in this transaction?

Actually, the star didn't need to know, because it didn't send out the photon long before the person was born. There are two main issues overlooked in the construction of this apparent paradox:

(1) In view of relativity, distances and time lapses are only relative, as is the time order of spacelike-separated events.
(2) No transaction can be set up without the availability of an absorber, in the present, so that any photon transfer establishes the emission event *in the past*.

First, consider point (1), with reference to Figure 4. Assume the star and the person (call him Bob) are in the same inertial frame. Event F is the star's demise. The story in which the star has ceased to exist before Bob comes along only holds relative to certain inertial frames. In fact, there is no invariant distance between the star and Bob, nor is there any invariant time of travel for the photon to get from the star to Bob. There is also no invariant time order of the events involving Bob's birth (denoted by B) and the star's emission event, since these are spacelike-separated. Consider a rocket ship traveling very fast from the star towards Bob. Its spatial axis (compressed to one dimension) is the slanted line intersecting point C. This means that point C is simultaneous with the star's emission according to the rocket's perspective. From Bob's perspective, he was born after the star emitted; but from the rocket's perspective, Bob was born before the star emitted.

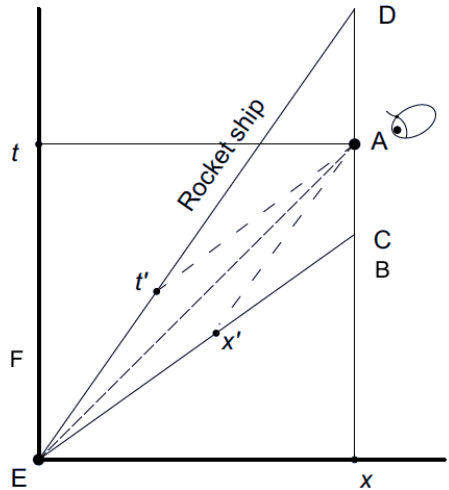

Figure 4. A star emits to a person who thinks that the star emitted and died before he was born.

According to the rocket, the spatial distance from the star to Bob is $x'$, and the time it takes for the photon to reach Bob is $t'$. These are much smaller than the values $x$ and $t$ assigned by Bob. Thus, from the rocket's perspective Bob is much closer to the star, and the time of the photon's travel is correspondingly reduced. In addition, from the standpoint of the rocket, the star dies only after Bob becomes available as an absorber (well after C, as can be seen by drawing a line parallel to the $x'$ axis from F to Bob's timeline). The lesson here is that the ability of a source to engage with absorbers is not restricted by sequences of events or spatiotemporal displacements relative to any particular inertial frame, since those are not absolute conditions. In this case, we see that according to the rocket ship, there is nothing perplexing about the star engaging in a transaction with Bob.

Regarding point (2): The advent of incipient transactions is governed by the absorbers in the present, and the actualized transaction acts to extrude the new spacetime interval from the present into the past, as a new element of the "spacetime fabric." In this sense, all transactions have a form of built-in retrocausation, but it is limited to the establishment of new spacetime events. It is not an influence contained within spacetime that affects or alters already-actualized events. Emitters and absorbers negotiate in the present (which we can identify with the quantum substratum) via offer waves (OW) and confirmation waves (CW), and it's only at the final stage of an actualized transaction that a "past event" is established corresponding to the actualized emission event. So, again, generation of OW and CW, which act in the quantum substratum as a form of *res potentia*, must be carefully distinguished from the actualized real photon that is a spacetime entity. The latter is a form of *res extensa*, as the connection between actualized emission and absorption events. This real photon is represented by a projection operator (outer product of the OW and CW components corresponding to the actualized transaction). Again, the actualized *events* that make up spacetime are *activities* of emitters and absorbers; the latter never become part of the spacetime manifold, remaining in the quantum substratum. In this sense, they are "eternally present."

## 7. Conclusion.

We have considered the manner in which the transactional process—a non-unitary quantum interaction—gives rise to an emergent spacetime manifold. As noted at the beginning of this paper, the idea that the spacetime manifold is emergent from the quantum level—as opposed to being an omnipresent "container" for all that is physically real—may seem radical, but arguably it is needed for full ontological consistency of the correspondence between the "fuzzy" quantum level, which seems to violate certain strictures of relativity, and the level of determinate spacetime events that unquestionably obeys relativity. We can gain insight into this matter (no pun intended) by considering the relationship of rest mass, as a source of the gravitational field, to the field itself. Note that the Einstein gravitational field equations are directly analogous to the Maxwell electromagnetic field equations, in the sense that the field is determined by its sources:

Maxwell equations (in covariant form):

$$\partial_\mu F^{\mu\nu} = \mu_0 J^\nu$$

Einstein equations (omitting cosmological constant $\Lambda$):

$$G_{\mu\nu} = \kappa T_{\mu\nu}$$

In both of these equations, the field generated by its sources is described on the left-hand side, while the sources are represented on the right-hand side. It should be noted that the electromagnetic field sources (charges, $J^\mu$) are not *in* the field. They are sources of the field, and as such, they are not contained within it: field lines terminate at the source charges. Analogously, the proper understanding with regard to gravitation is that the source of the gravitational field—matter—is not contained within the field, which is spacetime (more precisely, the geometric structure of spacetime). The gravitational field (i.e., spacetime and its structure) terminates at its sources, which are material systems. This again tells us that matter is not contained *within* spacetime, any more than charges are contained within the electromagnetic field they generate.

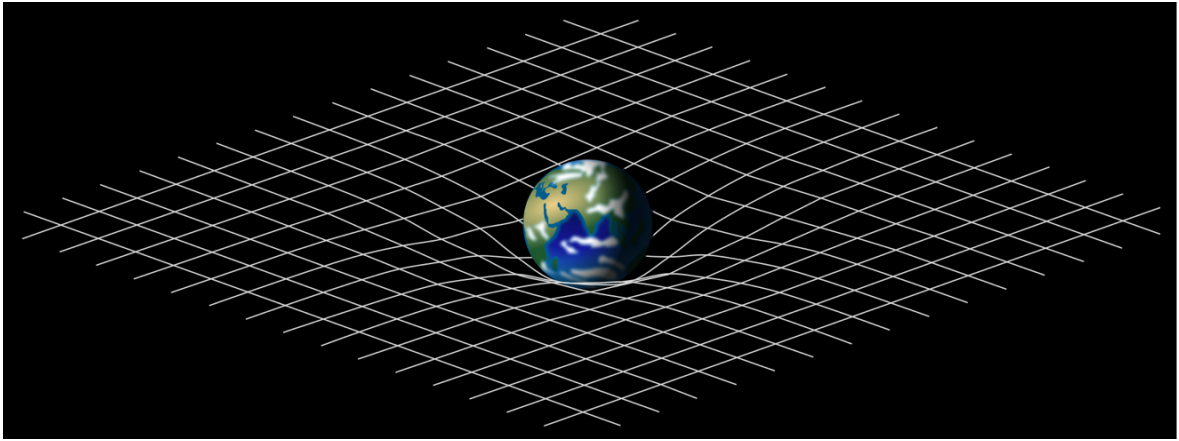

Figure 5. A matter source warps spacetime (here represented as a two-dimensional surface) from outside the spacetime manifold. Illustration courtesy of Wikimedia Commons.

Interestingly, illustrations of the warping of spacetime by matter by projecting the three spatial dimensions down to a 2-D plane depict matter sources outside the spacetime "fabric" (Figure 5). Despite the usual uncritical assumption that "everything physical is contained within spacetime," these representations illustrate (however unintentionally) the correct understanding: that matter sources give rise to the structure of spacetime from beyond it. More recently, with 3D animation programs, there have been attempts to depict matter sources curving spacetime in a three-dimensional depiction of the spacetime manifold while being contained within that same space. Yet even in those depictions, which try to place the matter sources "inside the spacetime container," the field structure terminates at the sources, implying that those sources are still outside the spacetime construct.

Einstein himself was uneasy about the relationship between the left-hand side of his equation (expressing the geometric structure of spacetime) and the right-hand side, containing the material energy-momentum tensor $T_{\mu\nu}$, a non-geometric object. As Paul Wesson and James Overduin put it:

> The geometrical object $G_{\mu\nu}$ is known as the Einstein tensor, and comprises the left-hand side of the field equations….However, it is not so widely known that Einstein wished to follow the same procedure for the other side of his field equations. That is, he wished to replace the common properties of matter, such as the density ρ and pressure p, by geometrical expressions. He termed the former "base wood" and the latter "fine marble." (Wesson and Overduin, 2019, p. 6)

According to the current proposal, the reason Einstein's goal of replacing matter by geometry was not achieved is because matter is the source of the spacetime structure, the latter being naturally described by its geometry. In contrast, that which *creates* the spacetime manifold is of a fundamentally different nature, just as electrical charges (electromagnetic field sources) are of a fundamentally different physical nature from the field to which they give rise, and they are neither part of it nor contained within it. As argued above, the material sources of the gravitational field (i.e., spacetime) are quantum systems, and as such are not contained within spacetime: quantum systems are physical possibilities, while spacetime is a structured manifold of actualities. Nevertheless, taking matter as beyond spacetime does not mean that it is deserving of the apparent contempt ("base wood") in which Einstein held it. On the contrary, arguably it is because of sophisticated and subtle topological and symmetry principles that matter can serve as the source of the spacetime manifold. For example, rest mass arises through fermionic spin, a topological property as noted in Section 4; and charge, which couples with the electromagnetic field and thereby gives rise to photon transfer and transactions resulting in spacetime events, can be understood in terms of a $U(1)$ gauge symmetry. None of these are properties of spacetime, but rather are properties of the physical possibilities—quantum systems—that are the sources of spacetime.

Acknowledgments. I am grateful to John Rather for bringing to my attention the work of Paul Wesson and colleagues.